\documentclass[twocolumn,showpacs,preprintnumbers,amsmath,amssymb]{revtex4}

\usepackage{graphicx}
\usepackage{dcolumn}
\usepackage{bm}

\begin{document}

\preprint{APS/123-QED}

\title{Miniature ceramic-anvil high-pressure cell for magnetic measurements in a commercial superconducting quantum interference device magnetometer\footnote{Review of Scientific Instruments {\bf{82}}, 053906 (2011).}}

\author{Naoyuki Tateiwa$^{1}$}
\author{Yoshinori Haga$^{1}$}%
\author{Zachary Fisk$^{1,2}$}
\author{Yoshichika {\=O}nuki$^{1,3}$} 
\affiliation{
$^{1}$Advanced Science Research Center, Japan Atomic Energy Agency, Tokai, Naka, Ibaraki 319-1195, Japan\\
$^{2}$ University of California, Irvine, California 92697, U.S.A\\
$^{3}$Graduate School of Science, Osaka University, Toyonaka, Osaka 560-0043, Japan
}%

\date{\today}

\begin{abstract}
A miniature opposed-anvil high-pressure cell has been developed for magnetic measurement in a commercial superconducting quantum interference device (SQUID) magnetometer. Non-magnetic anvils made of composite ceramic material were used to generate high-pressure with a Cu-Be gasket. We have examined anvils with different culet sizes (1.8, 1.6, 1.4, 1.2, 1.0, 0.8 and 0.6  mm). The pressure generated at low temperature was determined by the pressure dependence of the superconducting transition of lead (Pb). The maximum pressure $P_{max}$ depends on the culet size of the anvil: the values of $P_{max}$ are 2.4 and 7.6 GPa for 1.8 and 0.6 mm culet anvils, respectively.  We revealed that the composite ceramic anvil has potential to generate high pressure above 5 GPa. The background magnetization of the Cu-Be gasket is generally two orders of magnitude  smaller than the Ni-Cr-Al gasket for the indenter cell.  The present cell can be used not only with ferromagnetic and superconducting materials with large magnetization but also with antiferromagnetic compounds with smaller magnetization. The production cost of the present pressure cell is about one tenth of that of a diamond anvil cell. The anvil alignment mechanism is not necessary in the present pressure cell because of the strong fracture toughness (6.5 MPa${\cdot}$m$^{1/2}$) of the composite ceramic anvil. The simplified pressure cell is easy-to-use for researchers who are not familiar with high pressure technology. Representative results on the magnetization of superconducting MgB$_2$ and antiferromagnet CePd$_5$Al$_2$ are reported.

\end{abstract}

\maketitle

\section{Introduction}

Pressure is an important parameter in condensed matter physics~\cite{eremets}. Applying pressure changes lattice parameters, bonding angles, and overlap of the wave functions of electrons in a material. Consequently, the physical properties of the materials are changed and new physical phenomena can be expected at high-pressures. Many experimental and theoretical studies have been performed on pressure-induced phenomena such as the superconductor~\cite{buzea,shimizu1}.

  Magnetization is a fundamental physical quantity that characterizes the response of a material to applied magnetic field. Recently, it has become possible to measure the magnetization of a sample easily using a commercial superconducting quantum interference (SQUID) magnetometer. Several types of high-pressure cells have been made for use with these magnetometers. Most pressure cells designed for the commercial SQUID magnetometers are of piston-cylinder type~\cite{reich,diederichs,uwatoko1,kamishima, kamarad, kamenev,uwatoko2}. The piston-cylinder cell has adequate sample space ($\sim$10 mm$^3$) for precise magnetic measurements. But the maximum pressure is at most 1.5 GPa since the outer diameter of the pressure cell is limited to less than 9 mm for the commercial SQUID magnetometer. Opposed-anvil high-pressure cells such as diamond anvil (DAC) or indenter-type cells have been developed for commercial SQUID magnetometers~\cite{mito1,mito2,alireza,giriat,kobayashi}. Magnetic measurement can be done up to 15 GPa using the DAC. But the volume of the sample space in the DAC is less than 0.01 mm$^3$. It can be used for the ferromagnetic or superconducting compounds with large absolute magnetization below the transition temperature. Advanced techniques are necessary to use the DAC. The indenter-type cell is easier to use and the volume of the sample space is about 1 mm$^3$.  However, the background magnetization of the Ni-Cr-Al gasket in the indenter-type cell is quite large and the maximum pressure is about 2.5 $\sim$ 3.0 GPa~\cite{akazawa,hidaka,umeo}.

  We have studied the superconductivity and the magnetism in the rare earth and actinide compounds where many interesting physical phenomena have been discovered mainly in the pressure region from 2 to 5 GPa\cite{flouquet}.  The low background magnetization of the pressure cell is a necessary condition for a precise measurement.  In this study, a simple opposed-anvil high-pressure cell has been developed for magnetic measurements using the commercial SQUID magnetometer.  Non-magnetic anvils made of composite ceramic are used to generate high-pressure above 5 GPa. The background magnetization of the present pressure cell is significantly smaller than that of the indenter cell. The production cost of the present pressure cell is about one tenth of that of a diamond anvil cell. The present pressure cell without the anvil alignment mechanism is easy-to-use. Table I summarizes the features of several miniature high-preessure cells for magnetic measurements in the SQUID magnetometer. The details of the present ceramic anvil high-pressure cells are described in following sections.

    \begin{table*}[t]
\caption{\label{tab:table1}%
Miniature high-pressure cells for the commercial SQUID magnetometer. $P_{max}$ : maximum pressure, and $V_{sample}$: volume of sample space.}
\begin{ruledtabular}
\begin{tabular}{ccccc}
\textrm{high-pressure cell}&
\textrm{$P_{max}$ (GPa)}&
\textrm{$V_{sample}$ (mm$^3$)}&
\textrm{cost ($\$$)}&
\textrm{operation}\\
\colrule
piston cylinder-type cell\cite{diederichs,uwatoko1,kamishima,kamarad,kamenev,uwatoko2} & 1.0 $\sim$1.5 & $\sim$ 10 & $\sim$ 10$^3$&easy\\
indenter-type cell\cite{kobayashi} & 2.5 $\sim$ 3.0 & $\sim$ 1 & $\sim$ 10$^3$&easy\\
diamond anvil cell\cite{mito1,alireza,giriat} &$\sim$ 15 & $\sim$ 0.01& $\sim$ 10$^4$&difficult \\
\hline
present ceramic anvil cell & 2.4 $\sim$ 7.5 &0.015 $\sim$ 0.57 & $\sim$ 10$^3$&easy\\
\end{tabular}
\end{ruledtabular}
\end{table*}
        \begin{figure}[b]
\includegraphics[width=7.cm]{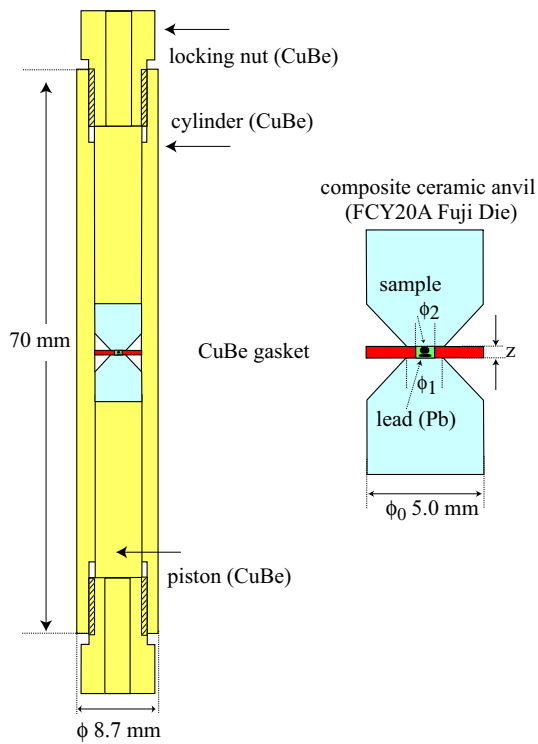}
\caption{\label{fig:epsart}(Color online) Cross-sectional views of the miniature high-pressure cell for the commercial SQUID magnetometer. The cell is 8.7 mm in diameter and 70 mm in length and is made up of two opposing ceramic anvils. The cell body, piston, locking nut and gasket are made of non-magnetic and hardened Cu-Be alloy.}
\end{figure}  

\section{high-pressure cell}
Figure 1 shows schematic drawings of the miniature high-pressure cell for use with a magnetic property measurement system (MPMS) from Quantum Design (USA)\cite{mpms}. The system is the most widely used SQUID magnetometer in the world and can resolve magnetic moment changes as small as 10$^{-8}$ emu over a wide range of temperatures and magnetic fields. The diameter of the inner bore of the MPMS is only 9 mm. The present pressure cell was manufactured by the R $\&$ D Support Co\cite{rd}. The designed pressure cell for this study is 8.7 mm in diameter and 70 mm in length. The cell can be introduced into the MPMS system in the same way one introduces a standard sample holder. The long cylinder is designed to reduce the contribution from cylinder and locknuts constructed out of a nonmagnetic hardened Cu-Be alloy (Japanese Industrial Standard No. C1720B-HT, 98$\%$Cu-2$\%$Be). A Cu-Be gasket separates two opposing composite ceramic anvils described below. The details of the composite ceramic anvil will be described below. There is no conventional anvil alignment mechanism in the cell and the anvil alignment is achieved through precision machining of the piston and cylinder. With this simplification, it becomes easier to assemble and use the present pressure cell. In the small DAC for the SQUID magnetometer, it is somewhat difficult to align the anvils correctly because of the smallness of anvil plate and screws for the tilt and $x$-$y$ adjustments.

   \begin{table}[b]
\caption{\label{tab:table1}%
Experimental conditions. ${\phi}_1$: culet size of the ceramic anvils,   ${\phi}_2$: diameter of the sample space,  $z$: initial thickness of the gasket before loading, and $P_{max}$ : maximum pressure.}
\begin{ruledtabular}
\begin{tabular}{cccc}
\textrm{Culet size ${\phi}_1$}&
\textrm{${\phi}_2$(mm)}&
\textrm{$z$ (mm)}&
\textrm{$P_{max}$ (GPa)}\\
\colrule
1.8 & 0.9 & 0.9 & 2.4 $\pm$ 0.1\\
1.6 & 0.8 & 0.8 & 2.5 $\pm$ 0.1 \\
1.4 & 0.7 & 0.7 & 2.8 $\pm$ 0.1 \\
1.2 & 0.6 &0.6 & 3.4 $\pm$ 0.3 \\
1.0 & 0.5 & 0.5 & 4.2 $\pm$ 0.2 \\
0.8 & 0.40 & 0.3 & 5.8 $\pm$ 0.4\\
0.6 & 0.30 &0.2 & 7.6 $\pm$ 0.4 \\
\end{tabular}
\end{ruledtabular}
\end{table}

 The culet sizes (${\phi}_1$) of the composite ceramic anvil are 0.6, 0.8, 1.0, 1.2, 1.4, 1.6, and 1.8 mm. Table II summarizes relations between the culet size of the anvil ${\phi}_1$ and the diameter of the sample space ${\phi}_2$, the initial thickness of the gasket before loading $z$ and the maximum pressure $P_{max}$. The outer diameter of the gasket and girdle one of the anvil are 5.0 mm. The cone angle of the anvil is 90$^\circ$.  As shown in the table II, the highest pressure of 7.6 GPa was achieved when the 0.6 mm culet anvils were used.

 A sample and pressure manometer lead (Pb) were loaded in the sample space filled with a pressure-transmitting medium Daphne 7373\cite{yokogawa,tateiwa}. The load was applied to the pressure cell through a piston and clamped by the locking nut at room temperature. The cell was installed in the MPMS magnetometer and cooled down to low temperatures.  In order to obtain a uniform and symmetrical signal in the SQUID coil system, dummy rods made of non-magnetic aluminum alloy were attached to both upper and lower sides of the pressure cell. The pressure at low temperatures were determined by the pressure dependence of the superconducting transition temperature in lead\cite{smith,eiling,wittig}.

  \begin{figure}[t]
\includegraphics[width=7.5cm]{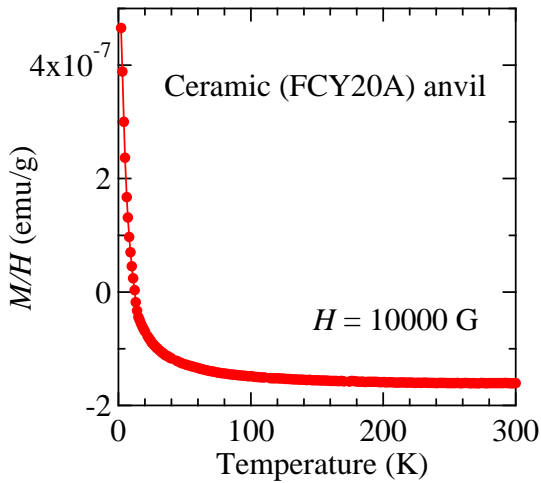}
\caption{\label{fig:epsart}(Color online)Temperature dependence of the magnetic susceptibility $M/H$ of the composite ceramic (FCY20A) anvil measured in a magnetic field of 10000 G. }
\end{figure} 
 We discuss the composite ceramic anvil. Usually, a tungsten carbide (WC) anvil is used in the opposed-anvil high-pressure cells such as the Drickamer and Bridgman type cells~\cite{eremets}. However, this is not appropriate for magnetic measurement since the background magnetization of the WC anvil is very large.  In this study, the cone-shaped anvils were made of non-magnetic composite ceramic (FCY20A) produced by the Fuji Die company, Japan\cite{fujidie}. The color of the ceramic is white. The composite ceramic is a mixture of Y$_2$O$_3$-partially stabilized zirconia (ZrO$_2$) and alumina (Al$_2$O$_3$) synthesized under high temperature and pressure. The hardness, compressive and tensile strengths of the ceramic are 1410 HV (Vickers), 4.12 and 1.86 GPa, respectively. The composite ceramics have the strong fracture toughness\cite{matsumoto}. The value of the fracture toughness for the present composite ceramic (FCY20A) is 6.5 MPa${\cdot}$m$^{1/2}$. The sapphire and cubic zirconia anvils have been used in the high pressure experiment over 10 GPa\cite{eremets,patterson,xu,russell,jackson}. These anvils, made of single crystals, are hard but brittle. The values of the fracture toughness for the materials are less than half of that for FCY20A\cite{wiederhorn,newcomb,cutler,tekeli}. 
The anvil alignment mechanism is necessary to avoid cracking of these anvils. Meanwhile, the mechanism is not required in the present pressure cell because of the stronger fracture toughness of the ceramic anvil. It has never been broken within our experiences.  Another merit of the ceramic anvil is its inexpensiveness: the price of the anvil per carat (0.2 g) is about two orders of magnitude smaller than that of the diamond anvil. The non-magnetic and non-electrically conductive ceramic anvil is appropriate for magnetic measurements under magnetic field. 
 
   Figure 2 shows the temperature dependence of the magnetic susceptibility in the ceramic anvil measured at 10000 G. In the high temperature region, $M/H$ is negative and nearly temperature-independent. The increase of $M/H$ at low temperatures can be ascribed to a small amount of impurities. The small value of the magnetic susceptibility is appropriate for magnetic measurement. The value of $M/H$ of the ceramic anvil is comparable to that of zirconia anvils and is two orders of magnitude smaller than that of WC anvils~\cite{uwatoko3}. 
      \begin{figure}[t]
\includegraphics[width=7.5cm]{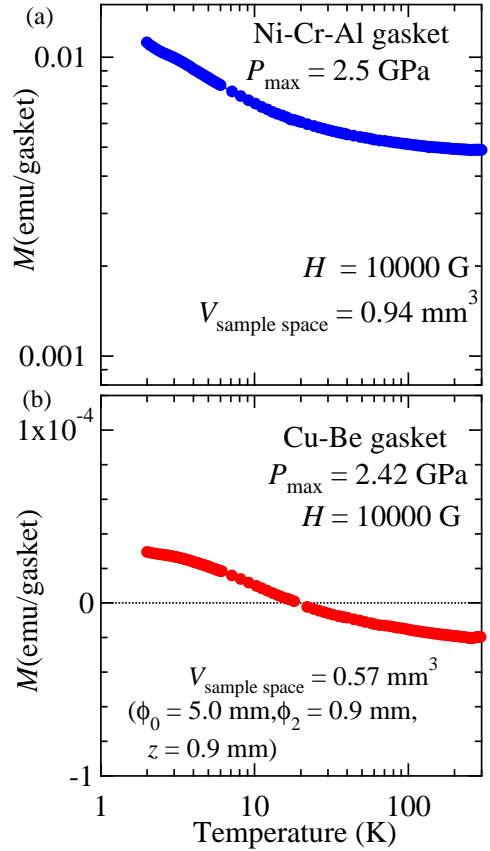}
\caption{\label{fig:epsart}(Color online)Temperature dependence of the magnetization in magnetic field of 10000 G for (a) the Ni-Cr-Al gasket in the indenter-type cell and (b) the Cu-Be gasket in the present miniature pressure cell. }
\end{figure} 

  We compare the background magnetization of the Cu-Be gasket in the present pressure cell with that of the Ni-Cr-Al gasket in the indenter cell.  We have prepared the indenter cell designed similarly to that in the reference 15. The outer and inner diameters, and the thickness of the Ni-Cr-Al gasket were 5.0, 1.2 and 0.9 mm, respectively. In the indenter-type cell, the gasket was pressed by the cone part of the zirconia anvil. The diameter of the anvil was 1.0 mm and the volume of the sample space was 0.94 mm$^3$. The maximum pressure of 2.50 GPa was achieved using the gasket.  The magnetization of the Ni-Cr-Al gasket under a magnetic field of 10000 G is shown in Figure 3 (a). The magnetization of the Cu-Be gasket for the present pressure cell is shown in Figure 3 (b). The outer and inner diameters, and the thickness of the Cu-Be gasket were 5.0, 0.9 and 0.9 mm, respectively. The volume of the sample space is 0.57 mm$^3$. The maximum pressure of 2.42 GPa was achieved using this gasket with the 1.8 mm culet anvil. The maximum pressures that can be attained using two gaskets are similar.  The magnetization of the Cu-Be gasket is about two orders of magnitude smaller than that of the Ni-Cr-Al gasket. Although the volume of the sample space in the Cu-Be gasket is about 60 $\%$ of that of the Ni-Cr-Al gasket, the ratio of the sample magnetization to that of the gasket is significantly larger in the present pressure cell. Due to the smaller background magnetization, it becomes possible to measure a sample with a small magnetization such as an antiferromagnet as shown later. 
The maximum pressure in the indenter cell could be increased if the Ni-Cr-Al gasket with the smaller sample space were pressed by the anvil with smaller culet size. However the ratio of the sample magnetization to the BG one from the gasket would become extremely smaller.
 

     \begin{figure}[t]
\includegraphics[width=8cm]{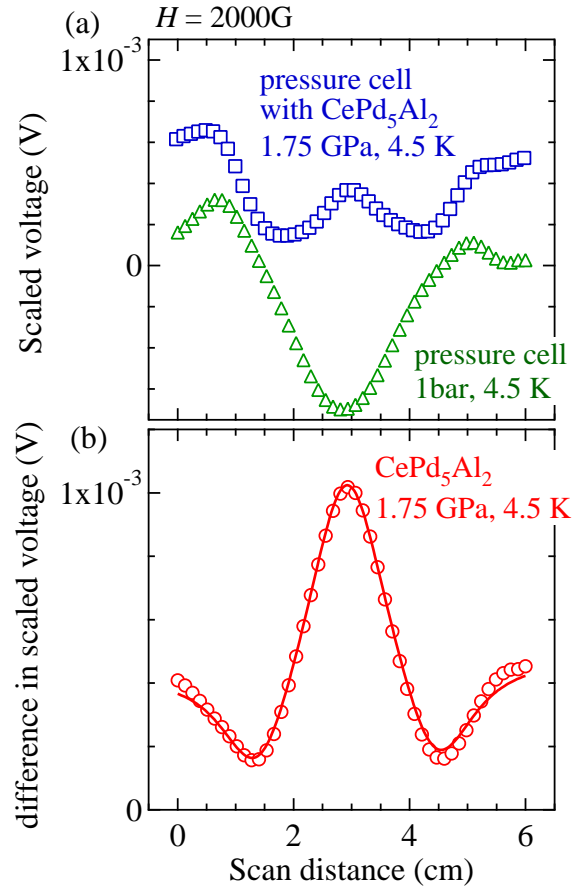}
\caption{\label{fig:epsart}(Color online)(a) Responses of the scaled SQUID voltage produced during the standard DC scan of length 6 cm. Open squares represent the response for the pressure cell with CePd$_5$Al$_2$ at 4.5 K and 1.75 GPa and open triangles does that for the empty pressure cell without the sample at 4.5 K and 1 bar. The measurements were done in a magnetic field of 2000 G. (b) Open circles represent the difference in the response with and without the sample. The line is the fit to the response using equation (1).}
\end{figure} 

   The magnetization of a sample in the pressure cell is obtained by subtraction of the magnetization of the empty pressure cell without the sample.  In the SQUID magnetometer MPMS, the magnetization is obtained automatically with MPMS software by fitting the SQUID response to a calculated form assuming a point dipole moment. This method is valid when the magnetization of the sample is large and the SQUID response is close to an ideal symmetrical form. However, when the magnetization of the sample is small and the SQUID response is not symmetrical, the fits by the software are misleading and can lead to false anomalies in the temperature or magnetic field dependences of the fitted magnetic moment. For accurate work, the SQUID response is collected with and without a sample and the difference signal is fitted to the calculated form using a specially written external program~\cite{subtraction}. Figure 4 (a) shows the responses of the scaled SQUID voltage produced during the standard DC scan of length 6 cm. The open squares represent the response of the pressure cell with an antiferromagnetic compound CePd$_5$Al$_2$ at 4.5 K and 1.75 GPa and the triangles that of the empty pressure cell without the sample at 4.5 K and 1 bar. The measurements were done in a magnetic field of 2000 G.  Figure 4 (b) shows the difference of the two responses with and without the sample.  The line is the fit to the response using a following equation (1)~\cite{subtraction}.

   \begin{eqnarray}
&&f(Z) = a_1+{a_2}{\cdot}Z\nonumber\\
&&+{a_3}{\cdot}{\Bigl[}{2\over {[R^2+(Z+a_4){^2}]^{3/2}}}-{1\over {[R^2+({\Lambda}+(Z+a_4)){^2}]^{3/2}}}\nonumber\\
&&-{1\over {[R^2+(-{\Lambda}+(Z+a_4)){^2}]^{3/2}}} {\Bigl]}
\end{eqnarray}
  

  Here, $a_1$, $a_2$, $a_3$, and $a_4$ are fitting parameters. ${\Lambda}$ and $R$ are the longitudinal coil separation and longitudinal radius of the coil in the MPMS SQUID magnetometer, respectively. The parameter $a_1$ is a constant offset voltage and the parameter $a_2$ takes into account a linear electronic SQUID drift during the data collection. The magnetization of the sample is obtained from the value of the parameter $a_3$ after multiplying by the apparatus parameters. The parameter $a_4$ is the shift of the sample along the axis of the magnet. The magnetization of CePd$_5$Al$_2$ obtained using this procedure is 3.19 $\times$ 10$^{-4}$ emu at 4.5 K and 1.75 GPa. The background magnetization of the cell was estimated as - 2.61 $\times$ 10$^{-4}$ emu, smaller than that of the sample at the same temperature. 
    
      In the next section, we will show experimental data on the superconductors lead (Pb) and MgB$_2$ and the antiferromagnet CePd$_5$Al$_2$. When superconductors are measured, the SQUID output response is generally close to the ideal form below the transition temperature because the large negative magnetization due to the Meissner effect is significantly larger than the magnetic back ground in low magnetic field. We will show the magnetization obtained by the MPMS software. The background magnetization was not subtracted. The magnetization of CePd$_5$Al$_2$ was small and it was obtained by the procedure as given above.

   \section{Measurements and Results}
  \subsection{Basic performance of the pressure cell}
 We measured the temperature dependence of Pb in a magnetic field of 10 G using anvils with several culet sizes as shown in Figure 5.  Experimental configurations such as the size of gasket for each anvil are given in Table II.  Experimental data using anvils with the culet sizes 1.8, 1.0, 0.8 and 0.6 mm are shown.  In these measurements, there is only a small piece of Pb in the sample space filled with the pressure medium Daphne 7373. At low temperatures, a large drop of the magnetization associated with the Meissner effect was observed. The pressure at low temperatures was determined from the pressure dependence of the superconducting transition temperature $T_{sc}$ of Pb\cite{smith,eiling,wittig}. The value of $T_{sc}$ was determined from a peak temperature in the temperature derivative of magnetization ${\partial M}/{\partial T}$ shown as arrows in the figure. As mentioned before, the absolute value of the negative magnetization of Pb in the superconducting state is large and the SQUID ouput is symmetrical. The magnetization shown in the figure is obtained from the analysis of the SQUID output by the MPMS software. The magnetization becomes negligibly small above $T_{sc}$. The superconducting transition temperature $T_{sc}$ of Pb is 7.19 K at ambient pressure in zero magnetic field. The transition temperature decreases with increasing pressure\cite{smith,eiling,wittig}. When the anvils with ${\phi}_1$ = 1.8 mm were used, the transition temperature is 6.3 K for an applied load of 946 kgf (0.946 ton). This indicates that the pressure inside the sample chamber is about 2.4 GPa. When the anvils with the smaller culet size are used, the maximum pressure becomes higher. The maximum pressure is 7.6 $\pm$ 0.4 GPa when the 0.6 mm culet anvils were used. It is noted that the error in the pressure is estimated from the transition width ${\Delta}T_{sc} $ of the superconducting transition.  
   \begin{figure}[t]
\includegraphics[width=7.5cm]{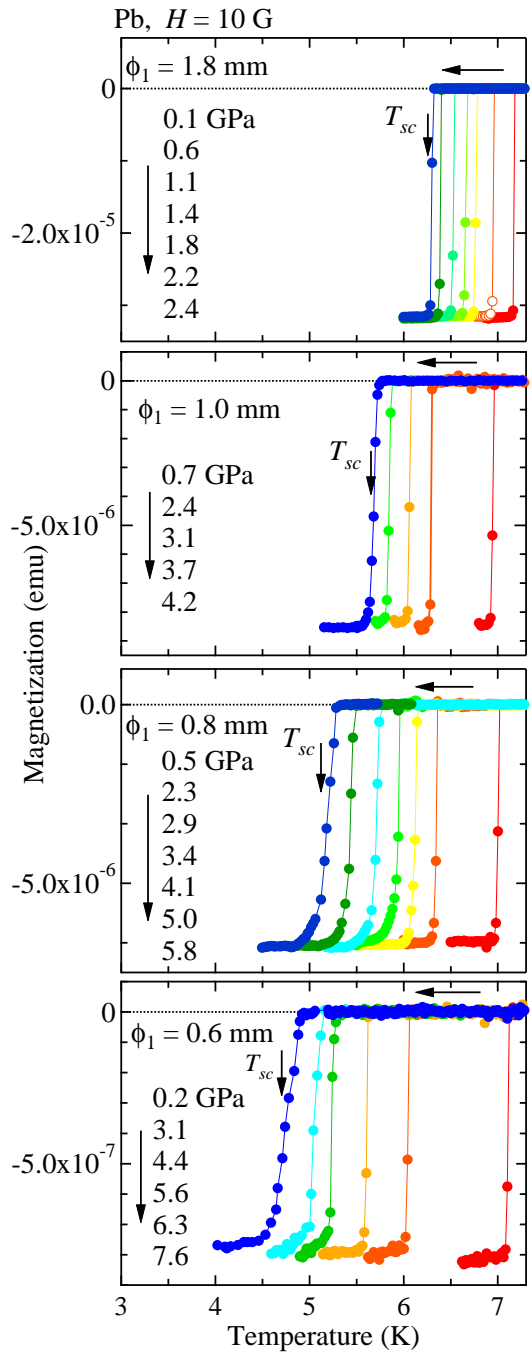}
\caption{\label{fig:epsart}(Color online) Temperature dependences of the magnetization of Pb in magnetic field of 10 G using several anvils with the culet sizes ${\phi}_1$ = 1.8, 1.0, 0.8 and 0.6 mm. }
\end{figure} 
    \begin{figure}[t]
\includegraphics[width=7.5cm]{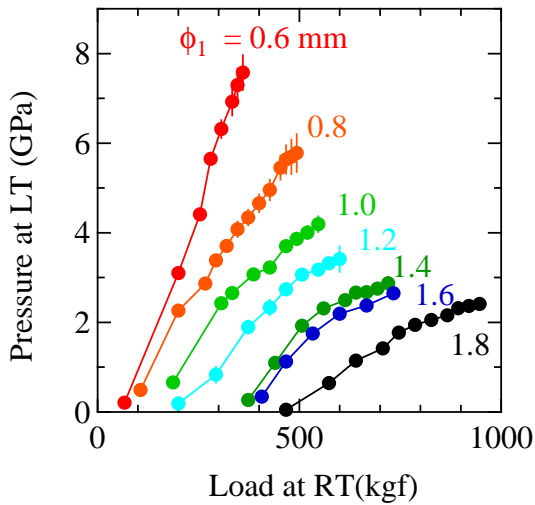}
\caption{\label{fig:epsart}(Color online) Relations between the applied forces at room temperature and the pressure values at low temperatures for anvils with ${\phi}_1$ = 1.8, 1.6, 1.4, 1.2, 1.0, 0.8 and 0.6 mm.}
\end{figure}

 The superconducting transition is very sharp in the lower pressure region and becomes broader in the higher pressure region. The solidification pressure of the pressure-medium Daphne 7373 is about 2.4 GPa\cite{yokogawa}. The broader transition in the higher pressure region may be due to the increase of the non-hydrostaticity in the pressure.
   
  Figure 6 shows relations between the applied loads at room temperature and the pressure values at low temperatures for the anvils with ${\phi}_1$ = 1.8, 1.6, 1.4, 1.2, 1.0, 0.8 and 0.6 mm. The pressure value increases with increasing load and shows a tendency to saturate at higher applied load. The maximum pressure and pressure efficiency becomes larger when the anvils with the smaller culet size were used. The maximum pressure is about 4.1 GPa for the anvil with ${\phi}_1$ = 1.0 mm.  We have not used the anvils with the culet size larger than 1.8 mm since the maximum load limit for the pressure cell is estimated as 1100 kgf (1.1 ton). 

         \begin{figure}[t]
\includegraphics[width=7.5cm]{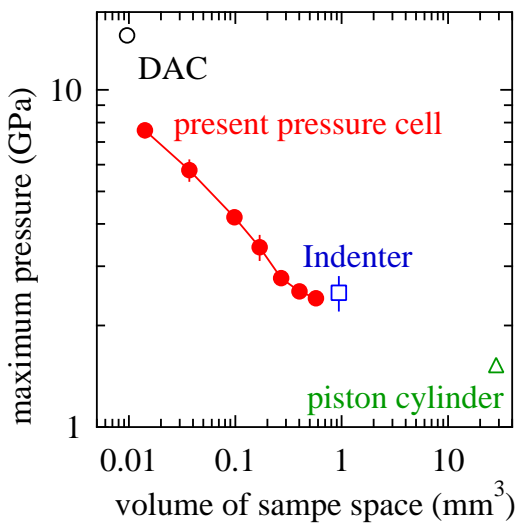}
\caption{\label{fig:epsart}(Color online) Relations between maximum pressure and volume of the sample space for the present pressure cell (closed circles) and indenter cell (open square) from our study, piston-cylinder cell (open triangle) and DAC (open circle). The data for the present cell and indenter cell are from our study, and those for the piston cylinder and DAC are from references 10 and 13, respectively.}
\end{figure} 

 The maximum pressures for the 0.8 and 0.6 mm culet anvils are 5.8 and 7.6 GPa, respectively. There values are larger than the compressive and tensile strengths of the composite ceramic FCY20A reported as 4.12 and 1.86 GPa, respectively\cite{fujidie}. The anvils are surrounded by the inner wall of the cylinder of the pressure-cell. We suppose that the massive support from the cylinder as well as the strong fracture toughness of the anvil is important to achieve higher pressures largely above the mechanical strength. 
 
    
 Figure 7 shows relations between maximum pressure and volume of the sample space in the present pressure cell (dotted circles), indenter cell (open square), piston-cylinder cell (open triangle) and DAC (open circle). The data for the present cell and indenter cell are from our study, and those for the piston cylinder and DAC are from the references 10 and 13, respectively. The volumes of the sample space in the present pressure cell are between those of the DAC and the indenter cell.  So far, the DAC has been the only pressure cell that allows dc magnetization measurement at pressures above 3 GPa in the commercial SQUID magnetometer. It is easier to do magnetic measurements with the present pressure cell than with the DAC even though the maximum pressure is limited to below 7.6 GPa. The production cost of the present pressure cell is about one tenth of that of the diamond anvil cell. Although the volume of the sample cell in the present cell is smaller than that in the indenter cell, the smaller background magnetization in the present pressure cell is advantageous for a reliable measurement.

\subsection{Examples of magnetization measurement under high-pressure}
 MgB$_2$ is a well-known superconductor with superconducting transition temperature $T_{sc}$ = 39 K at ambient pressure\cite{nagamatsu}. We studied the pressure effect of the superconducting transition temperature in the present pressure cell with the 0.6 mm culet anvils. The diameter of the sample space and the thickness of the Cu-Be gasket are 0.30 and 0.25 mm, respectively. A polycrystal sample of MgB$_2$ was placed in the sample space filled with the pressure-transmitting medium glycerin.  Figure 8 shows the temperature dependences of the magnetization in MgB$_2$ under magnetic field of 20 G at 1 bar, 3.9, 5.0 and 6.8 GPa. The data at 1 bar were obtained with the pressure cell where the load was not applied. The magnetization shows the sudden decrease due to the Meissner effect below the transition temperature $T_{sc}$ = 39 K at ambient pressure. With increasing pressure, the superconducting transition temperature decreases. The value of $T_{sc}$ is estimated as 32 K at 6.8 GPa. The pressure dependence of  $T_{sc}$ is roughly consistent with a previous study\cite{deemyad}. 
\begin{figure}[t]
\includegraphics[width=7.5cm]{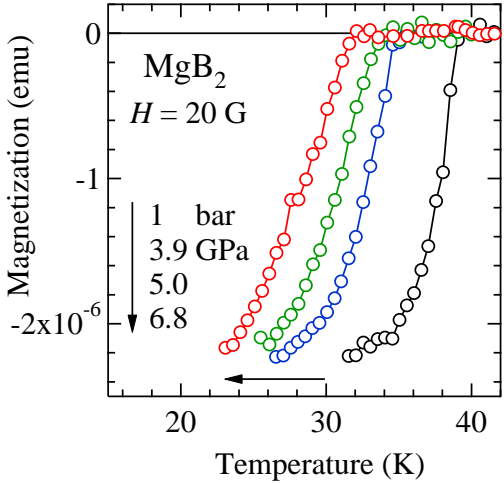}
\caption{\label{fig:epsart}(Color online)Temperature dependences of the magnetization of MgB$_2$ under magnetic field of 20 G at 1 bar, 3.9, 5.0, and 6.8 GPa. The data were obtained in the present pressure cell with the 0.6 mm culet anvils. }
\end{figure}  
        
      \begin{figure}[t]
\includegraphics[width=7.5cm]{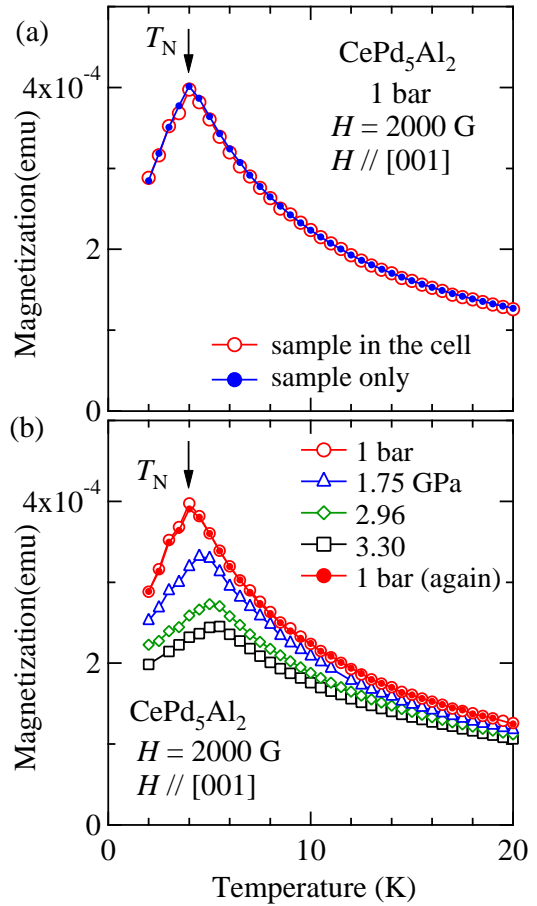}
\caption{\label{fig:epsart}(Color online) Temperature dependences of the magnetization of CePd$_5$Al$_2$ under magnetic field of 2000 G at (a)1 bar with and without using the pressure cell,  and (b) 1 bar, 1.75, 2.96,  and 3.30 GPa. The data were obtained by the present pressure cell using anvils with ${\phi}_1$ = 1.6 mm.}
\end{figure} 
 Next, we studied the pressure effect on the magnetic ordering temperature of antiferromagnet CePd$_5$Al$_2$ with antiferromagnetic transition temperature (N\'{e}el temperature) $T_{\rm N}$  = 4.1 K at ambient pressure\cite{nakano}. The pressure-induced superconductivity was observed above 10 GPa where $T_{\rm N}$ becomes 0 K\cite{honda}. The pressure effect on the antiferromagnetic transition temperature was studied up to 3.30 GPa with the present pressure cell using the 1.6 mm culet anvils. A single crystal sample was used for the high-pressure experiment. The magnetic field is applied along the magnetic easy axis (the [001] direction) in the orthorhombic crystal structure. 

  Since the magnetization of the antiferromagnet CePd$_5$Al$_2$ is small, the magnetization of the sample in the cell was obtained by the procedure with the equation (1) described before. The response of the SQUID voltage for the empty pressure cell without the sample was subtracted from that for the pressure cell with the sample. The difference between the responses was analyzed with the eq. (1). Open circles in Figure 9 (a) indicate the temperature dependence of the magnetization at 1 bar where the load was not applied to the cell. Closed circles indicate the magnetization data at 1 bar obtained by the measurement of only the sample in the commercial SQUID meter without using the pressure cell. The two data are coincident, indicating the reliability of the procedure for the determination of the magnetization.

 Figure 9 (b) shows the temperature dependences of the magnetization in CePd$_5$Al$_2$ at high-pressures in a magnetic field of 2000 G. The open circles, triangles, diamonds and squares are the magnetization data at 1 bar, 1.75, 2.96 and 3.30 GPa, respectively. At 1 bar, there is a peak in the temperature dependence at $T_{\rm N}$  = 4.0 K. With increasing pressure, the transition temperature shifts to higher temperatures. The value of $T_{\rm N}$ at 3.30 GPa is 5.5 K. This is consistent with the previous high-pressure study by the electrical resistivity measurement\cite{honda}. The closed circles in Fig. 9 (b) are the data at 1 bar after the high-pressure experiment. The measurement was done after the load had been released from the pressure cell. The data agrees with the one (open circles) taken before applying load to the pressure cell. Although the Cu-Be gasket becomes deformed during the loading process, the deformation of the gasket does not affect the background magnetization of the pressure cell. 

   \section{Conclusion}
We have developed a miniature opposed-anvil high-pressure cell for magnetic measurements in a commercial SQUID magnetometer. Non-magnetic anvils made of composite ceramic (FCY20A, Fuji Die Co.) were used with a Cu-Be gasket. Several anvils with different culet sizes (1.8, 1.6, 1.4, 1.2, 1.0, 0.8 and 0.6  mm) were tested. The maximum pressure $P_{max}$ depends on the culet size of the anvil: the values of $P_{max}$ are 2.4 and 7.6 GPa for the 1.8 and 0.6 mm culet anvils, respectively. Since the background magnetization of the Cu-Be gasket is smaller, the present cell is applicable not only to the ferromagnetic and superconducting materials with large magnetization but also to the antiferromagnetic compounds with smaller magnetization. The production cost of the present pressure cell is about one tenth of that of a diamond anvil cell. The simplified pressure cell is easy-to-use for researchers who are not familiar with high pressure technology. Experimental results for the magnetization measurements of superconducting MgB$_2$ and antiferromagnetic CePd$_5$Al$_2$ were reported.

   \section{Acknowledgments}
  This work was supported by a Grant-in-Aid for Scientific Research on Innovative Areas ``Heavy Electrons (No. 20102002, No. 23102726), Scientific Research S (No. 20224015), A(No. 23246174), C (No. 22540378), Specially Promoted Research (No. 20001004) and Osaka University Global COE program (G10) from the Ministry of Education, Culture, Sports, Science and Technology (MEXT) and Japan Society of the Promotion of Science (JSPS).

\bibliography{apssamp}

\end{document}